\documentclass[trackchanges, twocolumn]{aastex701}

\usepackage{graphicx}	% Including figure files
\usepackage{amsmath}	% Advanced maths commands

\newcommand{\msun}{{\,\rm M_\odot}}

\newcommand{\erg}{\,{\rm erg}}
\newcommand{\s}{\,{\rm s}}
\newcommand{\yr}{\,{\rm yr}}
\newcommand{\Gyr}{\,{\rm Gyr}}
\newcommand{\K}{\,{\rm K}}

\newcommand{\AU}{\,{\rm AU}}
\newcommand{\pc}{\,{\rm pc}}

\newcommand{\cMpc}{\,{\rm cMpc}}

\newcommand{\dex}{\,{\rm dex}}
\begin{document}

\title{Weighing Little Red Dots with Transient Events}

\author[orcid=0009-0003-6068-6921,sname='Tran']{Vinh Tran}
\affiliation{Department of Physics, California Institute of Technology, Pasadena, CA 91125, USA}
\affiliation{Department of Physics and Kavli Institute for Astrophysics and Space Research, Massachusetts Institute of Technology, Cambridge, MA 02139, USA}
\email[show]{vinhtran@caltech.edu}  

\author[orcid=0000-0002-6196-823X,sname='Shen']{Xuejian Shen}
\affiliation{Center for Astrophysics, Harvard and Smithsonian, 60 Garden Street, Cambridge, MA 02138, USA}
\affiliation{Department of Physics and Kavli Institute for Astrophysics and Space Research, Massachusetts Institute of Technology, Cambridge, MA 02139, USA}
\email{xuejianshen@fas.harvard.edu}  

\author[orcid=0000-0003-1811-8915,sname='Zier']{Oliver Zier}
\affiliation{Center for Astrophysics, Harvard and Smithsonian, 60 Garden Street, Cambridge, MA 02138, USA}
\email{oliver.zier@cfa.harvard.edu}

\author[orcid=0000-0002-2380-9801]{Anna de Graaff}\thanks{Clay Fellow}
\affiliation{Center for Astrophysics, Harvard and Smithsonian, 60 Garden Street, Cambridge, MA 02138, USA}
\affiliation{Max-Planck-Institut f\"ur Astronomie, K\"onigstuhl 17, D-69117 Heidelberg, Germany}
\email{degraaff@mpia.de}

\author[orcid=0000-0003-3997-5705,sname='Naidu']{Rohan P. Naidu}
\affiliation{Institute for Astronomy, University of Hawaii, 2680 Woodlawn Drive, Honolulu, HI 96822, USA}
\email{rnaidu@hawaii.edu}

\author[orcid=0000-0001-8593-7692,sname='Vogelsberger']{Mark Vogelsberger}
\affiliation{Department of Physics and Kavli Institute for Astrophysics and Space Research, Massachusetts Institute of Technology, Cambridge, MA 02139, USA}
\email{mvogelsb@mit.edu}

\begin{abstract}
Recent JWST observations have revealed a large population of compact red sources at $z \gtrsim 4$, known as Little Red Dots (LRDs), many of which show signatures of accreting massive black holes (BHs). The physical nature of these sources and their connection to host galaxies are under debate. We propose an independent avenue for constraining their nature through transient phenomena, such as tidal disruption events (TDEs) and quasi-periodic eruptions (QPEs), arising from interactions between a star and the gas envelope surrounding the BH. These event rates depend sensitively on BH mass and provide a way to ``weigh'' LRDs. We calculate the expected TDE and QPE rates in LRDs under three distinct scenarios: (1) LRDs are truly overmassive BHs, (2) LRDs have BH masses following the classical local scaling relations (and the reported BH masses in observations are overestimated), and (3) the currently observed LRDs are only the tip of the iceberg of a larger population of low-mass BHs. We find that the predicted TDE and QPE rates differ dramatically across scenarios, especially in the presence of steep stellar cusps. The expected TDE rates per degree-square, assuming a Hernquist stellar distribution with a Bahcall--Wolf cusp embedded, are $2.78 \times 10^{-3}$, $1.96 \times 10^{-3}$, and $3.37 \times 10^{-2} \yr^{-1} \deg^{-2}$ for the three scenarios, respectively, while the QPE rates are $1.64 \times 10^{-2}$, $4.72 \times 10^{-2}$, and $4.96 \times 10^{-1} \yr^{-1} \deg^{-2}$. Upcoming wide-field surveys with Euclid, Roman, and LSST may be capable of detecting these high-redshift transient events and obtaining light curves, which encode additional information about the BH mass and the gas structure of LRDs. Stellar transient events will provide valuable insight into the early assembly of massive BHs.
\end{abstract}



\section{Introduction}
\label{sec:intro}

Recent JWST observations have revealed a surprisingly abundant population of compact, red sources at $z\gtrsim 4$, the so-called Little Red Dots (LRDs), which do not map cleanly onto the familiar categories of galaxies or classical active galactic nuclei \citep[AGN; e.g.,][]{Matthee2024,Greene2024,Kokorev2024,Kocevski2025,Akins2025}. Observationally, LRDs are characterized by a blue rest-frame UV slope and a rapidly rising red optical continuum, compact or unresolved (optical) morphologies, and, in many cases, broad Balmer emission lines as signatures of accreting massive black holes~\citep[BHs; e.g.,][]{Setton2025,Hviding2025}. Meanwhile, LRDs are unusually faint in X-ray given their optical luminosities \citep{Yue2024,Ananna2024,Maiolino2025}, often show weak mid-IR and far-IR emission \citep[e.g.,][]{Setton2025a,Casey2025,Akins2025,Xiao2025}, and exhibit little or no continuum variability \citep[at least in the majority; e.g.,][but see also \citealt{Zhang2025}]{Hayes2024,Kokubo2025,Tee2025,Zhang2025a,Liu2026}, challenging to explain by simple dust-reddening of a standard Type I AGN. A particularly important clue is the presence of Balmer breaks~\citep[e.g.,][]{Ji2025,Naidu2025,DeGraaff2025} in at least a subset of LRDs, some of which are too strong to be explained with a stellar origin.

Recent spectroscopic observations have pushed the field toward a more complicated picture. At least some LRDs host accreting BHs embedded in extremely dense gas cocoons~\citep[e.g.,][]{Naidu2025,DeGraaff2025,Inayoshi2025,Rusakov2026}. These new findings strongly affect BH mass estimates. Many published masses rely on single-epoch virial scalings based on broad Balmer line widths and continuum luminosities, which are highly uncertain for LRDs. The geometry of the broad-line region and the corresponding virial factors are poorly constrained, while electron scattering in dense gas can broaden the Balmer lines and bias virial mass estimates~\citep[e.g.,][]{Yue2024,Rusakov2026,Greene2026}. A recent dynamical measurement~\citep{Juodzbalis2025} provides an important cross-check in at least one lensed LRD, although the population may be heterogeneous. The face-value BH masses of LRDs, and in general low-luminosity broad-line AGN (BLAGN) found by JWST, lie systematically above the local scaling relations between BH mass and their host galaxy stellar mass~\citep[e.g.,][]{Pacucci2023,Maiolino2024,Durodola2025}, although this is debated given the aforementioned uncertainties. The picture may be further complicated by selection effects, since current LRD samples across the literature sometimes do not have fully consistent selection criteria and may be subjected to contamination and incompleteness~\citep[e.g.,][]{Hainline2025,Kocevski2025,Akins2025}. This has led to speculation that LRDs, and in general BLAGN observed by JWST, may appear overly massive because of selection effects~\citep[e.g.,][]{Li2025}. Establishing the true correlation between these low-mass early BHs and their host galaxy properties is vital for deciphering their dominant seeding and growth channels~\citep[e.g.,][]{Volonteri2010,Greene2020,Regan2024,Shen2025}.

These uncertainties motivate independent ways of constraining the BH masses and duty cycles of LRDs. Transient phenomena associated with BHs, such as tidal disruption events (TDEs) or quasi-periodic eruptions (QPEs), may provide a complementary probe. For example, TDEs occur when a star is torn apart by the tidal forces of a BH, after which a fraction of the stellar debris remains bound and accretes, producing a luminous flare~\citep[e.g.,][]{Hills1975,Rees1988,Evans1989,Phinney1989}. Since the first detections~\citep{Bade1996,Komossa1999a,Komossa1999b,Greiner2000}, multiwavelength observations over the past two decades have established TDEs as a distinct class of nuclear transients spanning the optical/UV, X-ray, and sometimes radio bands~\citep[e.g.,][]{Komossa2015,Gezari2021}. Since TDE rates depend sensitively on BH mass, they offer a promising probe of the BH mass function and occupation fraction in the low- and intermediate-mass regimes, which are otherwise difficult to access directly~\citep[e.g.,][]{WangMerritt2004, Stone2016, Yao2023}. Further constraints on BH masses can be obtained with the lightcurves~\citep{Yao2023}. The prospects for detecting TDEs at $z\gtrsim 4$ are now becoming increasingly compelling. Recently, \citet{Karmen2025} reported HZTDE-1 in COSMOS-Web, a promising candidate at $z\sim 5$, although a superluminous supernova interpretation has not yet been excluded. Looking ahead, the next generation time domain surveys, with e.g. JWST and the Vera C. Rubin Observatory Legacy Survey of Space and Time (LSST), will be extremely powerful in searching for TDEs at high redshifts. tTDEs in LRD hosts and descendants offer a potentially independent handle on whether LRDs host genuinely overmassive BHs or instead represent the tip of the iceberg of a larger population of lower-mass BHs~\citep{Inayoshi2024,KarChowdhury2024,Karmen2026}. Moving beyond TDEs, many LRDs exhibit optical-to-near-IR spectral energy distributions resembling stellar photospheres~\citep[e.g.][]{Naidu2025,DeGraaff2025}. The inferred photospheric radii can be much larger than the tidal disruption radius, raising the possibility of a different class of transients: instead of being tidally disrupted directly by the BH, an incoming star may first collide with, plunge through, or be engulfed by the extended gaseous envelope surrounding the BH \citep{Suzuguchi2026}. Such an interaction could produce a QPE in LRDs and provide complementary constraints on the density, geometry, and dynamical structure of the material powering LRD emission.

The paper is structured as follows: Section \ref{sec:loss_cone} details the loss cone dynamics of the BH--stellar system considered in our analysis and the stellar distribution and population assumed for LRDs. Section \ref{sec:result} presents the resulting per-galaxy TDE and QPE rates, as well as the corresponding per-degree-squared rates, under three different LRD population assumptions. Finally, Section \ref{sec:discussion} concludes the paper with a discussion of the results. All code used in our analysis is publicly available\footnote{\url{https://github.com/vinh-qtran/TidalDisruptionLRD}}.

% In this section, we calculate both the per-galaxy and all-sky TDE rates for LRDs\footnote{\url{https://github.com/vinh-qtran/TidalDisruptionLRD}}. Section \ref{subsec:rho_star} describes the stellar distributions considered in this study. Section \ref{sec:loss_cone} presents the loss-cone dynamics and the calculation of per-galaxy TDE rates. Section \ref{subsec:stellar_mf} discusses the stellar mass function assumed for LRDs. Finally, Section \ref{subsec:all_sky_rate} computes the all-sky TDE rate for these galaxies.

\section{Methods}
\label{sec:loss_cone}

\subsection{The loss cone}
\label{ssec:loss_cone_condition}

We first consider the TDE of a star by a naked BH. We follow \cite{WangMerritt2004} to compute the TDE rates. For an BH of mass $M_\bullet$, stars that pass within the tidal radius
\begin{align}
    \label{eqn:disrupt_radius}
    r_{\rm t} & = R_\ast \left(\eta^2 \frac{M_\bullet}{m_\ast}\right)^{1/3} \\
    & = 0.415 \AU \, \left(\frac{R_\ast}{{\rm R}_\odot}\right) \left(\frac{m_\ast}{{\rm M}_\odot}\right)^{-1/3} \left(\frac{M_\bullet}{10^6 \, {\rm M}_\odot}\right)^{1/3}
\end{align}
are tidally disrupted and subsequently accreted. Here, $m_\ast$ and $R_\ast$ denote the stellar mass and radius, respectively, and $\eta = 0.844$ for an $n=3$ polytropic stellar structure. This tidal disruption condition is equivalent to the angular momentum limit of
\begin{equation}
    \label{eqn:loss_cone_angular_momentum}
    J^2 \leq J_{{\rm{lc}}}^2 \equiv 2 r_{\rm t}^2 \left(\psi(r_{\rm t}) - \epsilon\right),
\end{equation}
where $\epsilon \equiv \psi - v^2/2$ denotes the stellar binding energy, with $\psi$ and $v$ being the negative gravitational potential and star particle speed, respectively. This is often referred to as the consumption loss cone.

This assumes the tidal radius satisfies $r_{\rm t} > 2 r_{\rm g} = 2 G M_\bullet / c^2$; otherwise, the star is swallowed whole rather than being tidally disrupted. This condition imposes an upper limit on the BH mass that produces TDEs, known as the Hills mass,
\begin{equation}
    \label{eqn:Hill_mass}
    M_{\rm H} = 9.67 \times 10^7 \msun \left( \frac{R_\ast}{{\rm R}_\odot}\right)^{3/2} \left( \frac{m_\ast}{{\rm M}_\odot}\right)^{-1/2}.
\end{equation}
For solar-type stars, $M_{\rm H}$ lies at the upper end of the reported BH mass range in LRDs. We enforce this constraint in our calculations of TDE rates. In addition, we consider the QPE scenario, in which the BH is surrounded by a dense gas envelope with a characteristic photosphere radius of $r_{\rm ph} \sim 1000 \AU$ \citep[e.g.][]{deGraaff2025_2,Kido2025,Sun2026,Umeda2026,Suzuguchi2026}. This is several orders of magnitude larger than the tidal disruption radius, which typically has the value of $r_{\rm t} \lesssim 1 \AU$. Our calculation of the QPE rate follows that of TDE, with the only modification being the replacement of $r_{\rm t}$ by $r_{\rm ph} = 1000 \AU$. We note that, because $r_{\rm ph} \gg r_{\rm t}$, the regime in which $\epsilon \sim \psi(r_{\rm ph})$, where the calculation either becomes invalid or is dominated by numerical errors, is significantly enlarged. As later discussed, this is especially true when $r_{\rm ph}$ approaches within $\sim 1 \text{--} 2 \dex$ of the BH influence radius $r_{\rm h}$, defined as the radius within which the enclosed stellar mass equals the BH mass, i.e. $M_\ast(r_{\rm h}) = M_\bullet$. It must also be noted that this calculation of the QPE rate represents only the birth rate of QPE systems, as a star whose orbit crosses within the photosphere radius may produce repeating events thereafter.

%In some cases, the photosphere radius can be comparable to the BH influence radius $r_{\rm h}$, defined as the radius within which the enclosed stellar mass equals the BH mass, i.e. $M_\ast(r_{\rm h}) = M_\bullet$. This, as later shown, suppresses most of the flux and drastically reduces the disruption rate.

\subsection{Diffusion of stars into the loss cone}
\label{ssec:diff}

Assuming no mass segregation, the stellar distribution function follows
\begin{equation}
    \label{eqn:eddington_eqn}
    f(\epsilon) = \frac{1}{\sqrt{8}\,\pi^2\langle m_\ast \rangle} \frac{{\rm d}}{{\rm d} \epsilon} \int_0^\epsilon \frac{{\rm d} \rho}{{\rm d} \psi} \frac{{\rm d} \psi}{\sqrt{\epsilon - \psi}},
\end{equation}
with $\rho$ as the stellar density profile. Here, $\langle m_\ast^n \rangle = \int \left({\rm d}N_\ast / {\rm d} m_\ast\right) m_\ast^n \, {\rm d} m_\ast$, where ${\rm d}N_\ast / {\rm d} m_\ast$ denotes the stellar mass function.
In addition to assuming spherical symmetry, we adopt an isotropic velocity distribution and require the system to be gravitationally bound, such that $f(\epsilon \leq 0) = 0$. From this distribution function, the local angular momentum diffusion coefficient can then be computed following \cite{MagorrianTremaine1999}
\begin{multline}
    \label{eqn:local_diff_coeff}
    \lim_{R \to 0} \frac{\langle(\Delta R)^2\rangle}{2R} = \frac{32 \pi^2 r^2 G^2 \langle m_\ast^2 \rangle \ln{\Lambda}}{3 J_{\rm c}^2 (\epsilon)} \\
    \times \left(3 I_{1/2} (\epsilon) - I_{3/2} (\epsilon) + 2 I_{0} (\epsilon)\right).
\end{multline}
Here, $R \equiv J^2 / J_{\rm c}^2$, where $J_{\rm c}(\epsilon)$ denotes the angular momentum of a circular orbit with energy $\epsilon$. We adopt $\Lambda = 0.4 \, M_\bullet / \langle m_\ast \rangle$ following \cite{SpitzerHart1971}. The terms within the parentheses represent moments of the distribution function
\begin{align}
    \label{eqn:I_0}
    I_0 (\epsilon) &= \int_0^\epsilon f(\epsilon^\prime) \, {\rm d}\epsilon^\prime \\
    \label{eqn:I_1/2}
    I_{n/2} (\epsilon) &= \left[2 \left(\psi(r) - \epsilon\right)\right]^{-n/2} \notag \\
    & \hspace{1cm} \times \int_\epsilon^{\psi(r)} \left[2 \left(\psi(r) - \epsilon^\prime\right)\right]^{n/2} f(\epsilon^\prime) \, {\rm d}\epsilon^\prime.
\end{align}
The orbit-averaged angular momentum diffusion coefficient is
\begin{equation}
    \label{eqn:orbit_diff_coeff}
    \bar{\mu} (\epsilon) = \frac{2}{P(\epsilon)} \int_0^{r(\epsilon)} \frac{{\rm d} r^\prime}{v_r (r^\prime)} \lim_{R \to 0} \frac{\langle(\Delta R^2)\rangle}{2R},
\end{equation}
with the orbital period $P(\epsilon) = 2 \int_0^{r(\epsilon)} {\rm d}r^\prime / v_r (r^\prime)$ calculated from the radial velocity of $v_r(r) = \left[2 \left(\psi(r) - \epsilon\right)\right]^{1/2}$.

The flux of stars per unit time and per unit energy into the loss cone then follows \cite{CohnKulsrud1978}
\begin{equation}
    \label{eqn:flux_loss_cone}
    \mathcal{F} (\epsilon) = 4 \pi^2 J_{\rm c}^2(\epsilon) P(\epsilon) \bar{\mu}(\epsilon)\frac{f(\epsilon)}{\ln R_0^{-1}},
\end{equation}
where $R_0(\epsilon)$ denotes the angular momentum threshold below which no stars remain. In general, $R_0(\epsilon)$ can be smaller than $R_{\rm lc}(\epsilon)$ and takes the form
\begin{equation}
    \label{eqn:R_0}
    R_0 (\epsilon) = R_{\rm lc} (\epsilon) \begin{cases}
        \exp{(-q)} & [q \geq 1] \\
        \exp{(-0.186\,q - 0.824\,\sqrt{q})} & [q < 1]
    \end{cases}
\end{equation}
with $q (\epsilon) = P (\epsilon) \bar{\mu} (\epsilon) / R_{\rm lc} (\epsilon)$ being the ratio of the orbital period to the loss cone filling timescale. $q \gg 1$ represents the ``full loss cone'' (or ``pinhole'') regime, where the loss cone is refilled much faster than the tidal disruption rate.

Calculating the loss cone flux in terms of dimensionless profiles $\tilde{r}, \tilde{\rho}, \tilde{\psi},$ and $\tilde{\epsilon}$ following Appendix \ref{apd:dimless}, using the units of mass, length, and velocity of $[M] = M_\bullet$, $[r] = r_{\rm h}$, and $[v] = \sqrt{G M_\bullet / r_{\rm h}}$, we obtain the ratio $q (\tilde{\epsilon})$ and loss cone flux $\mathcal{F} (\tilde{\epsilon})$
\begin{align}
    \label{eqn:q_ast}
    q (\tilde{\epsilon}) &= \frac{32 \pi^2}{3 \sqrt{2}} \ln{(\Lambda)} \frac{\langle m_\ast^2 \rangle}{M_\bullet \langle m_\ast \rangle} \left(\frac{r_{\rm t}}{r_{\rm h}}\right)^{-2} \frac{\tilde{h} (\tilde{\epsilon})}{\tilde{\psi} (r_{\rm t} / r_{\rm h}) - \tilde{\epsilon}}, \\
    \label{eqn:F_ast}
    \mathcal{F} (\tilde{\epsilon}) &= \frac{256 \pi^4}{3 \sqrt{2}} \frac{\ln{(\Lambda)}}{\ln{R_0^{-1}}} \frac{\langle m_\ast^2 \rangle}{\langle m_\ast \rangle^2} \sqrt{\frac{G M_\bullet}{r_{\rm h}^3}} \tilde{h} (\tilde{\epsilon}) \, \tilde{g} (\tilde{\epsilon}).
\end{align}
Here, $\tilde{g} (\tilde{\epsilon})$ and $\tilde{h} (\tilde{\epsilon})$ are the dimensionless form of $f (\epsilon)$ and the orbital average of $3 I_{1/2} (\epsilon) - I_{3/2} (\epsilon) + 2 I_{0} (\epsilon)$, respectively (Equation \ref{eqn:g_ast} and \ref{eqn:h_ast}).

The total per-galaxy TDE rate would then be
\begin{equation}
    \label{eqn:tde_rate}
    \dot{N}_{\rm TDE} = \int \mathcal{F} (\tilde{\epsilon}) {\rm d} \tilde{\epsilon}.
\end{equation}
As $q(\tilde{\epsilon})$ and $\mathcal{F}(\tilde{\epsilon})$ depend on the stellar mass and radius via their dependence on $r_{\rm t}$, the TDE rate per galaxy must then be obtained by integrating over the distributions of these stellar properties.

In most cases, $\mathcal{F}(\epsilon)$ exhibits a strong peak around $\epsilon \simeq \epsilon_{\rm h} \equiv \psi(r_{\rm h})$. The approximate calculation of the TDE rate for a density profile $\rho \sim r^{-\gamma}$, based on the behavior of $\mathcal{F}(\epsilon)$ and $R_0(\epsilon)$ near $\epsilon_{\rm h}$ \citep{WangMerritt2004}, yields
\begin{equation}
    \label{eqn:TDE_approx}
    \dot{N}_{\rm TDE} \sim {M_\bullet}^{\delta}; \hspace{0.5cm} \delta = \frac{27 - 19 \gamma}{6 \left(4 - \gamma\right)}.
\end{equation}
This predicts that, for $\gamma < 1.42$, the TDE rate increases with increasing BH mass, consistent with the calculated TDE rates for systems with $\gamma \leq 1.2$ presented in \cite{WangMerritt2004}. This is significantly different from the approximation of \cite{Alexander2017}, where $\dot{N}_{\rm TDE} \sim N_{\rm h} / t_{\rm rel,h}$. Here, $N_{\rm h} \sim M_\bullet / m_\ast$ is the total number of stars within $r_{\rm h}$, and $t_{\rm rel,h} \sim \left(P_{\rm h} / N_{\rm h}\right)\left(M_\bullet / m_\ast\right)^2$ is the relaxation time, with $P_{\rm h} \sim r_{\rm h}^{3/2} / M_\bullet^{1/2}$ denoting the orbital period at $r_{\rm h}$. This yields $\delta = \left(2 - 6/\gamma\right)^{-1}$, which is always negative for $\gamma < 3$, a behavior not observed in our full calculations presented later in Section~\ref{ssec:rate}.

We also note that, when $r_{\rm t}$ approaches within $\sim 1 \text{--} 2 \dex$ of $r_{\rm h}$ and $\psi(r_{\rm t}) \sim \epsilon_{\rm h}$ (as in the case of $r_{\rm t} = r_{\rm ph}$), the integration of $\mathcal{F}(\epsilon)$ will miss a non-negligible fraction at $\epsilon > \psi(r_{\rm t})$ and the flux near $\psi(r_{\rm t})$ can become dominated by numerical errors. The latter results in an artificial bump in the flux $\mathcal{F}(\epsilon)$, which in some cases becomes comparable to the flux around $\epsilon_{\rm h}$. When this occurs, we set the corresponding event rate to that of the lowest BH mass where the calculation remains reliable, and report this as the lower limit of the event rate. This is justified by the fact that the QPE rate increases monotonically with decreasing BH mass, as discussed in Section \ref{ssec:rate}.

\subsection{Stellar distribution}
\label{ssec:star}

Canonically, TDE rate calculations for the local galaxy population rely on two-dimensional fits to the surface brightness profile $I(R)$ to infer the spatial distribution of stars $\rho(r)$ \citep[e.g.][]{MagorrianTremaine1999,Syer1999,WangMerritt2004,Stone2016}. Assuming spherical symmetry and a constant mass-to-light ratio, the deprojection is achieved via the Abel inversion
\begin{equation}
    \label{eqn:Abel_inversion}
    \rho (r) \propto - \int_r^\infty \frac{{\rm d} I}{{\rm d} R} \frac{{\rm d} R}{\sqrt{R^2 - r^2}}.
\end{equation}
In these calculations, $I(R)$ is typically parameterized by the ``Nuker'' profile \citep{Lauer1995}, a double power law that behaves as $I(R) \sim R^{-\Gamma}$ at small $R$. An alternative choice is the core-S\'ersic profile \citep{Graham2003,Trujillo2004}, which matches a S\'ersic profile \citep{Sersic1963} at large $R$ onto an inner power law, and exhibits the same small-$R$ behavior. The nuclear star cluster (NSC) is usually not modeled as a separate component in these studies. The Nuker fits adopted by \cite{Stone2016} in fact explicitly excluded nuclear light excesses \citep{Stone2016b}. This simplification is reasonable for the massive early-type galaxies that dominate these samples, where the nucleation fraction is low, and the loss-cone flux originates mainly from stars near the influence radius $r_{\rm h}$ (as discussed in the previous section), well beyond the few-parsec extent of typical NSCs in these systems \citep{Neumayer2020}. In low-mass galaxies, by contrast, the sizes of NSCs become comparable to $r_{\rm h}$ and can enhance the TDE rate by up to two orders of magnitude \citep{Stone2016b,Pfister2020,Hannah2024}. 

However, for LRDs, the stellar distribution is highly uncertain. Recent decompositions of LRD spectra into a compact LRD core and a host galaxy \citep{Sun2026,Ishikawa2026,Umeda2026} motivate a natural starting point: the size--mass relation of normal star-forming galaxies at the same redshift. Some LRDs indeed show extended rest-UV emission, with half-light radii of $\sim 200\text{--}300 \pc$ \citep{Cloonan2026}, comparable to the $\sim 400 \pc$ rest-UV sizes of star-forming galaxies at $z = 5\text{--}7$ \citep{Morishita2024}. Yet this extended light need not trace stars. For example, it can be nebular emission from gas photoionized by the compact central source, and it is often offset from the LRD itself \citep[e.g.,][]{Rinaldi2024,Chen2025,Cloonan2026,Ji2026,Baggen2026}. Therefore, we consider the scenario in which the stellar component of LRDs is more compact than that of normal star-forming galaxies at the same redshift. A number-density argument sets the compactness. The abundance of LRDs is $\sim 2\dex$ below that of the galaxy population at similar UV luminosity \citep[e.g.][]{Greene2024,Kokorev2024,Kocevski2025}. If LRD hosts occupy the most compact wing of the size--mass distribution, they should sit roughly $2-3\sigma$ below the median relation. Stacked rest-optical imaging of LRDs supports this picture: the extended host component is $\sim 2.5$ times smaller than star-forming galaxies of similar stellar mass \citep{ZhangY2025}. As an estimate, we adopt the rest-optical size--mass relation of normal galaxies at $z \approx 5$ \citep{Miller2025},
\begin{equation}
    \label{eqn:size_mass_func}
    R_{\rm eff} = R_0 \, \left(M_\ast/10^{8} \msun\right)^{0.27}.
\end{equation}
We take $R_{\rm eff}$ as a proxy for the projected half-mass radius. For normal galaxies, $R_0 = 507 \pc$. For LRDs, if we shift $R_0$ downward by $2$ ($3$) times the inferred intrinsic scatter therein ($\sigma = 0.24\dex$), we obtain $R_0 = 168 \pc$ ($97 \pc$). At $M_\ast = 10^{7}\text{--}10^{8} \msun$, the characteristic LRD host mass \citep[e.g.,][]{Matthee2025,Sun2026}, the resulting $R_{\rm eff}$ is comparable to the $\lesssim 100 \pc$ upper limits on the rest-optical sizes of LRDs themselves \citep{Baggen2024,Guia2024,Cloonan2026}.

This is unlikely to be the end of the story. A significant fraction of LRDs remain unresolved, with $R_{50} \lesssim 100 \pc$ from the rest-UV through the rest-optical \citep{Cloonan2026}. Strongly lensed systems push the constraints further down to a few tens of parsecs, where substructures are revealed along with the red compact source \citep{Furtak2023,Furtak2024-lense,Baggen2026,Golubchik2026-VENUS,Yanagisawa2026-VENUS}. These observations allow for a dense nuclear stellar structure in at least part of the population, which would produce a more compact and cuspy density profile than a single S\'ersic component. Motivated by this, we construct a density profile for the host of the LRDs, which bridges the Dehnen profile \citep{Dehnen1993} (the galaxy component) and a Bahcall--Wolf cusp \citep{BahcallWolf1976} within $r_{\rm h}$ (where two-body relaxation drives the stellar distribution toward $\rho \propto r^{-7/4}$),
\begin{multline}
    \label{eqn:dehnen_cusp}
    \rho (r) = \rho_\ast \left(\frac{r}{r_{\rm h}}\right)^{\gamma-7/4} \left[1 + \left(\frac{r}{r_{\rm h}}\right)^{\beta}\right]^{-\frac{\gamma-7/4}{\beta}} \\
    \times \left(\frac{r}{a}\right)^{-\gamma} \left(1 + \frac{r}{a}\right)^{\gamma - 4}.
\end{multline}
Here, $\rho_\ast$ is the normalizing density, such that the total stellar mass is $M_\ast$, and $a$ is the scale radius. The last two factors constitute a Dehnen profile with inner slope $\gamma = \Gamma + 1$, the deprojected counterpart of the surface brightness slope $\Gamma$. In the composite model, $\gamma$ sets the logarithmic density slope in the intermediate region between $r_{\rm h}$ and $a$. For a pure Dehnen profile without the cusp, $\gamma = 0$ corresponds to a cored profile, while $\gamma = 1$ and $\gamma = 2$ recover the Hernquist \citep{Hernquist1990} and Jaffe \citep{Jaffe1983} profiles, respectively. The first two factors impose $\rho \propto r^{-7/4}$ at $r \ll r_{\rm h}$ and reduce to unity at $r \gg r_{\rm h}$, with $\beta$ controlling the sharpness of the transition. We choose $\beta = 4$ as a balance between sufficient sharpness and the numerical stability of subsequent calculations. The scale radius $a$ follows from $R_{\rm eff}$ through the $\gamma$- and $r_{\rm h}$-dependent projected half-mass radius of the density profile. Because the model already contains a dense nuclear cusp, the extended envelope implied by the size--mass relation is not necessarily inconsistent with the compact rest-UV sizes reported for individual LRDs. The lensing constraints apply to the compact, light-dominating components, whereas a low surface brightness envelope can evade detection against the glare of the nucleus \citep{Baggen2026}.

\begin{figure}
    \centering
    \includegraphics[width=\linewidth]{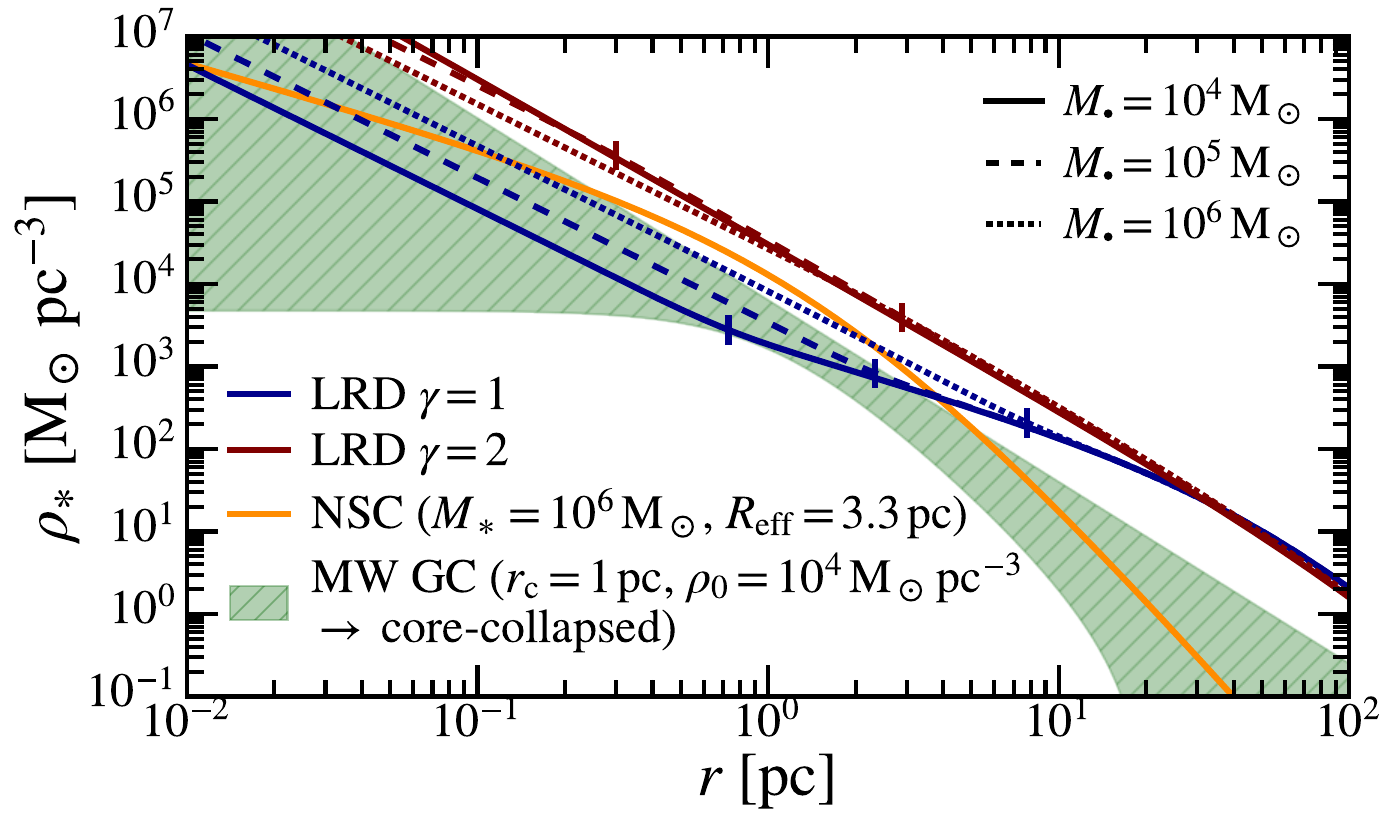}
    \caption{The stellar density profiles of LRDs (Equation \ref{eqn:dehnen_cusp}) assuming a stellar mass of $M_\ast = 10^8\msun$ and an effective radius of $R_{\rm eff} = 168\pc$. A wide range of BH masses, $M_\bullet = 10^4, 10^5, 10^6$, is shown in solid, dashed, and dotted lines, respectively, for two choices of $\gamma = 1$ (blue) and $\gamma = 2$ (red). For comparison, the density profile of a typical NSC in a galaxy of the same stellar mass (orange) is shown, along with the density range of typical higher-mass MW GCs to core-collapse specimens (green band). The inner regions of LRDs have densities comparable to those of NSCs and MW GCs for $\gamma = 1$, while for $\gamma = 2$, LRDs appear to be $\sim 0.5\dex$ denser than the core-collapsed GC.}
    \label{fig:density}
\end{figure}

\begin{figure}
    \centering
    \includegraphics[width=\linewidth]{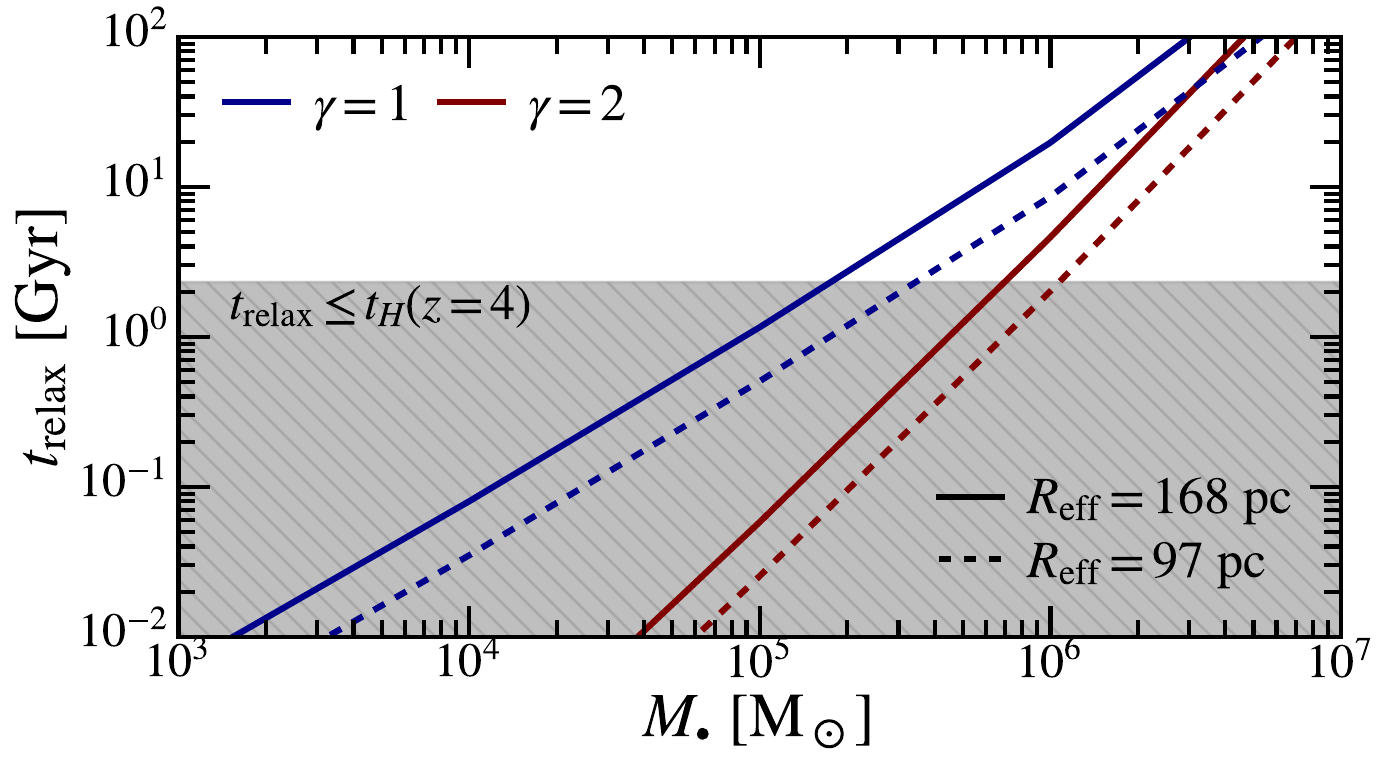}
    \caption{The relaxation times $t_{\rm relax}$ of LRD cusps as a function of BH mass $M_\bullet$, assuming a stellar mass of $M_\ast = 10^8\msun$ and effective radii of $R_{\rm eff} = 168\pc$ (solid) and $R_{\rm eff} = 97\pc$ (dashed). Calculations for the intermediate slopes of $\gamma = 1$ and $\gamma = 2$ are shown in blue and red, respectively. The relaxation times are compared to the Hubble time at $z = 4$ $t_H(z=4)$ (grey fill). As a result of their higher densities, the Bahcall--Wolf cusps in the $\gamma = 2$ case can reach equilibrium within the available time even for $M_\bullet = 10^6\msun$, whereas for $\gamma = 1$, only systems with BH masses below $10^5\msun$ can fully establish the cusp.}
    \label{fig:t_relax}
\end{figure}

Figure \ref{fig:density} shows the density profiles given by Equation \ref{eqn:dehnen_cusp} for a stellar mass of $M_\ast = 10^8\msun$ and an effective radius of $R_{\rm eff} = 168\pc$, following Equation \ref{eqn:size_mass_func}. We consider three BH masses, $M_\bullet = 10^4, 10^5,$ and $10^6\msun$, and two different values of the intermediate slope, $\gamma = 1$ and $\gamma = 2$. For comparison, we show the density profiles of several known compact systems in the Universe. The first is an NSC in a host galaxy of the same stellar mass $10^8\msun$. The NSC has the Hernquist profile with mass $10^6\msun$ and effective radius $3.3\pc$, according to the scaling relations in \citet{Neumayer2020}. Additionally, we show the density profiles of Milky Way (MW) globular clusters (GCs), spanning from a typical higher-mass GC, modeled with the King profile \citep{King1962} with $\rho_0 = 10^4 \msun \pc^{-3}$ and $r_{\rm c} = 1\pc$ \citep[e.g.][]{Harris1996,MeylanHeggie1997}, to a core-collapsed GC, modeled as $\rho = \rho_0 (r / r_{\rm c})^{-2.2}$ with $\rho_0 = 10^6 \msun \pc^{-3}$ and $r_{\rm c} = 0.1\pc$, similar to the values inferred for the core-collapsed, potentially BH-hosting GC M15 \citep[e.g.][]{Lauer1991,Gerssen2002}. Overall, the LRD cusps have densities comparable to those of NSCs and MW GCs for $\gamma = 1$, while in the denser $\gamma = 2$ case, the LRD densities are $\sim 0.5\dex$ higher than these compact systems. The BH mass has little effect on the density profiles in the latter case, as the slope of the Bahcall--Wolf cusp is similar to $\gamma = 2$.

An implicit assumption we made about the density profile is that the inner cusp has reached an equilibrium (the Bahcall--Wolf solution) within the life time of the LRD hosts. This relaxation time is roughly the two-body relaxation time of the cusp \citep{Bar-Or2013},
\begin{equation}
    \label{eqn}
    t_{\rm relax} = 0.077 \frac{\sigma^3}{G^2 \langle m_\ast \rangle \rho \ln{\Lambda}}.
\end{equation}
This definition is $4.4$ times smaller than the commonly used relaxation time that characterizes the diffusion timescale of the kinetic energy \citep{BinneyTremaine2008}. Here, $\rho$ is evaluated at the influence radius $r_{\rm h}$, while $\sigma$ is approximated as $\sigma^2 = GM_\bullet / r_{\rm h}$. For simplicity, we take the average mass of stars as $\langle m_\ast \rangle = 1\msun$, which results in $\Lambda = 0.4\, M_\bullet / {\rm M}_\odot$. In Figure~\ref{fig:t_relax}, we show $t_{\rm relax}$ of an LRD host with stellar mass $M_\ast = 10^8\msun$, and compare it against the Hubble time at $z = 4$ as a proxy for the age of the Universe $t_{H}(z=4) = 2.29\Gyr$, assuming the \citet{Planck2018} cosmology. We find that for $\gamma = 2$, even systems hosting massive $\sim 10^6\msun$ BHs can reach equilibrium by $z=4$. However, for $\gamma = 1$, where the density is significantly lower, systems with massive BHs may have difficulty fully relaxing to the Bahcall--Wolf cusp within the available time, even when the effective radius is reduced to $R_{\rm eff} = 97\pc$, corresponding to $3\sigma$ below the typical size of a high-redshift galaxy. In systems hosting overmassive BHs, discussed later in Section~\ref{ssec:rate}, the inner cusp may therefore not have reached a steady state. This points to a broader limitation of our treatment, which is entirely static. As the LRD and the associated nuclear star cluster form and evolve, the transient phenomena it produces, and their rates, need not resemble the equilibrium values adopted here. Our framework does not cover this assembly phase, nor the possibility that some systems never reach equilibrium at all. Capturing these effects requires a self-consistent theoretical framework for the seeding and evolution of early BHs, which we defer to future work.

% In addition,we consider the classical isothermal profile as a reference
% \begin{equation}
%     \label{eqn:iso}
%     \rho_{\rm Iso} (r) = \frac{\sigma^2}{2 \pi G r^2}.
%     \end{equation}
% Here, $\sigma$ is the stellar velocity dispersion, which we calculate using the $M_\bullet\text{--}\sigma$ relation calibrated by \cite{KormendyHo2013}
% \begin{equation}
%     \label{eqn:Mbh_sigma_relation}
%     M_\bullet = 3.09 \times 10^8 \msun \, \left(\frac{\sigma}{200 \kms}\right)^{4.38}.
% \end{equation}
% Note that for the isothermal profile, the total stellar mass is not well defined. Other $M_\bullet\text{--}\sigma$ relations could also be adopted; nevertheless, the results remain comparable between cases. For simplicity, we focus on this specific calibration. We emphasize that these relations are derived for the local galaxy population rather than LRDs, and thus may not accurately represent the properties of LRDs. For this profile, the influence radius is given by $r_{\rm h,Iso} = G M_\bullet / 2 \sigma^2$.

The last piece is the assumptions about the stellar population. Following \cite{Stone2016}, we adopt the stellar present-day mass function (PDMF) evolved from the Kroupa initial mass function (IMF). The corresponding PDMFs are given by
% \begin{equation}
%     \label{eqn:Salpeter_pdmf}
%     \chi_{\rm Sal} = \frac{{\rm d} N_\ast}{{\rm d} m_\ast} \bigg|_{\rm Sal} = \begin{cases}
%         A_{\rm Sal} \, \left(m_\ast / M_\odot\right)^{-2.35} &, m_\ast^{\rm min} < m_\ast < m_\ast^{\rm max} \\
%         0  &, \text{otherwise}
%     \end{cases}
% \end{equation}
% and
\begin{equation}
    \label{eqn:Kroupa_pdmf}
    \frac{{\rm d} N_\ast}{{\rm d} m_\ast} \bigg|_{\rm Kro} = \begin{cases}
        A \, \left(m_\ast / {\rm M}_\odot\right)^{-1.3} & [m_\ast^{\rm min} < m_\ast < 0.5 \, {\rm M}_\odot] \\
        0.5 \, A \, \left(m_\ast / {\rm M}_\odot\right)^{-2.3} & [0.5 \, {\rm M}_\odot < m_\ast < m_\ast^{\rm max}] \\
        0  & [\text{otherwise}],
    \end{cases}
\end{equation}
where we adopt $m_\ast^{\rm min} = 0.08 \msun$ and $m_\ast^{\rm max} = 2 \msun$, which represent a stellar population of age $T \sim 1 \Gyr$ (roughly the age of the Universe at $z\simeq 6$). Here, $A$ is the normalizing factor depending on the values of $m_\ast^{\rm min}$ and $m_\ast^{\rm max}$. We assume the relation $R_\ast \propto m_\ast^{\alpha}$, with $\alpha \simeq 0.8$ for lower main-sequence stars and $\alpha \simeq 0.57$ for their higher-mass counterparts. The boundary is set at $m_{\rm mid} = 1.66 \msun$ \citep{DemircanKahraman1991}. The tidal radius is then written as
\begin{equation}
    \label{eqn:r_t_ast}
    r_{\rm t} = \eta^{2/3} \, R_\odot \,\left(\frac{m_{\rm mid}}{{\rm M}_\odot}\right)^{0.8} \left(\frac{m_\ast}{M_\bullet}\right)^{\alpha - 1/3} \left(\frac{M_\bullet}{m_{\rm mid}}\right)^{\alpha}
\end{equation}
This implies that the TDE rate depends strictly on the stellar mass, and therefore on the stellar PDMF. Varying $m_\ast^{\rm max}$ over the range of $1 \msun$ to $10 \msun$ changes the TDE rate by less than a factor of $2$, which is much smaller than the effect of varying $\gamma$, which can alter the TDE rate by orders of magnitude. Other species, specifically the stellar remnants, such as white dwarfs, neutron stars, and stellar-mass BHs, can also contribute to the overall TDE rate. However, as noted by \cite{Stone2016}, the populations of white dwarfs and neutron stars are typically small, with individual masses comparable to those of main-sequence stars, limiting their overall impact. In contrast, stellar-mass BHs can provide a more significant contribution. In particular, for cusp profiles with higher BH masses or in core galaxies, such as in some of the cases considered here, the inclusion of stellar-mass BHs can enhance the TDE rate by a factor of $\gtrsim 1.5$. Determining the exact values proves challenging as stellar metallicity distribution and mass segregation for the LRD population remain uncertain. As a result, we exclude the species in our current analysis.

\section{Results}
\label{sec:result}

\begin{figure}
    \centering
    \includegraphics[width=\linewidth]{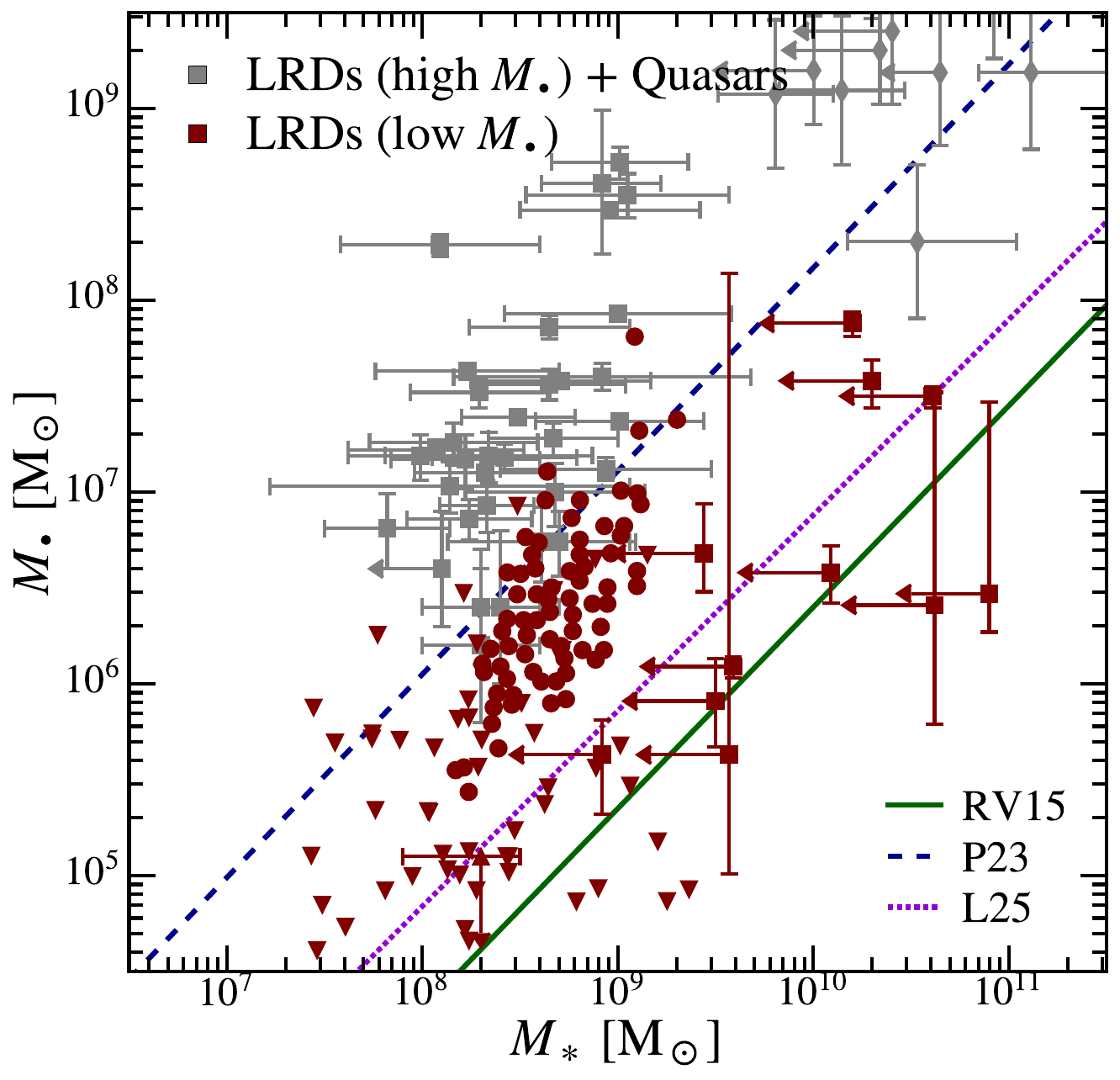}
    \caption{The BH mass $M_\bullet$ and total stellar mass $M_\ast$ of LRDs. LRDs with reported overmassive BHs compiled from \cite{Harikane2023,Maiolino2024,Juodzbalis2024,Kocevski2023,Kocevski2025,Taylor2025,PerezGonzalez2026}, as well as quasars from \citep{Stone2024,Yue2024,Ding2023}, are shown in grey. LRDs with more moderate BH masses are shown in red. Square markers indicate BH mass measurements based on the virial method, with those in the lower-$M_\bullet$ group calibrated for electron scattering \citep{Rusakov2026}. Diamonds mark quasars, while triangles indicate measurements based on the escape velocity argument \citep{Naidu2026}. Circles and inverted triangles indicate the BH envelope model with $\lambda_{\rm Edd} = 0.5$ \citep{Umeda2026} and the quasi-star (QS) model with $M_\bullet / M_{\rm QS} = 0.36$ \citep[the middle value of][]{Gentile2026}, respectively. The $M_\bullet \text{--} M_\ast$ relations corresponding to the local galaxy population \citep[RV15,][]{ReinesVolonteri2015}, observed LRDs \citep[P23,][]{Pacucci2023}, and the inferred intrinsic relations that account for hidden BLAGN \citep[L25,][]{Li2025} are shown as solid, dashed, and dotted lines, respectively.}
    \label{fig:relations}
\end{figure}

To compute the per-degree square TDE and QPE rate, we integrate the per-galaxy rate over the distribution of LRDs at $z = 4 \text{--} 6$
\begin{equation}
    \label{eqn:all_sky_rate}
    \dot{\mathcal{N}} = \frac{1}{\Omega_{\rm sky}} \int_{z=4}^{z=6} \dfrac{{\rm d} V_{\rm c}(z)}{{\rm d}z} \, n_{\rm LRD} \, \frac{\langle \dot{N} \rangle}{1 + z} \,{\rm d}z.
\end{equation}
Here, $\Omega_{\rm sky} = 41252.96 \deg^2$ is the total solid angle of the sky in degree square. ${\rm d} V_{\rm c}(z)$ is the comoving shell volume between $z$ and $z + {\rm d}z$, and $n_{\rm LRD}$ is the comoving number density of LRDs obtained by integrating the bolometric luminosity functions in \cite{Greene2026} (with updated bolometric corrections for LRDs, see also \cite{Shen2026-lumina}) within the range of $L_{\rm bol} = 10^{42.5} \text{--} 10^{45.5} \erg \s^{-1}$. The total comoving volume within $z = 4 \text{--} 6$ using Planck cosmology \citep{Planck2018} is $V_{{\rm c}} \simeq 8.5 \times 10^{11} \cMpc^3$, while $n_{\rm LRD} \simeq 1.2 \times 10^{-4} \cMpc^{-3}$. The factor of $1 / (1 + z)$ accounts for time dilation between the rest and observed frames. $\langle \dot{N} \rangle$ is the rest-frame average per-galaxy TDE rate, assuming a uniform distribution in stellar mass within the range $M_\ast = 10^{7} \text{--} 10^{9} \msun$, the median of which is motivated by \cite{Matthee2025,Lin2026}. Assuming a Gaussian distribution centering at $M_\ast = 10^{8} \msun$ with a standard deviation of $1 \dex$ results in no significant difference. We consider three scenarios:
\begin{enumerate}
    \item All LRDs are assumed to follow the $M_\bullet\text{--}M_\ast$ relation from \cite{Pacucci2023}, hereafter P23, which is calibrated using $M_\bullet$ (via virial kinetics method) and $M_\ast$ measurements from \cite{Harikane2023} and \cite{Maiolino2024}. \cite{Jones2025} revisited this relation using a more recent dataset \citep[e.g.][]{Juodzbalis2024,Kocevski2023,Kocevski2025,Taylor2025,PerezGonzalez2026}, along with a stricter definition of LRDs \citep{Kocevski2025}, and found consistent results.
    \item All LRDs instead follow the local $M_\bullet\text{--}M_\ast$ relation of \cite{ReinesVolonteri2015}. This relation is $\sim 1.5 \dex$ lower than the P23 relation. This scenario is supported by the recalibration of the bolometric luminosity presented in \cite{Greene2026}, as well as by an alternative measurement approach proposed in \cite{Rusakov2026, Naidu2026, Umeda2026, Gentile2026}.
    \item Finally, following \cite{Li2025}, hereafter L25, the observed LRDs and their virial-inferred MB masses represent only a small subset of the intrinsic population, which more closely follows the local-universe relation. In this scenario, only $\sim 9 \%$ of the total LRD population is observed due to selection effects and statistical uncertainties.
\end{enumerate}
The $1 \sigma$ scatters in the $M_\bullet\text{--}M_\ast$ relations, $\sigma = 0.69, 0.55,$ and $0.97 \dex$ for the three relations, respectively, are assumed to follow a Gaussian distribution in logarithmic space. All observational data for LRDs, as well as for a subset of quasars \citep{Stone2024,Yue2024,Ding2023}, along with the scaling relations considered here, are shown in Figure~\ref{fig:relations}.

\subsection{The TDE \& QPE rates}
\label{ssec:rate}

\begin{figure*}
    \centering
    \includegraphics[width= 1 \linewidth]{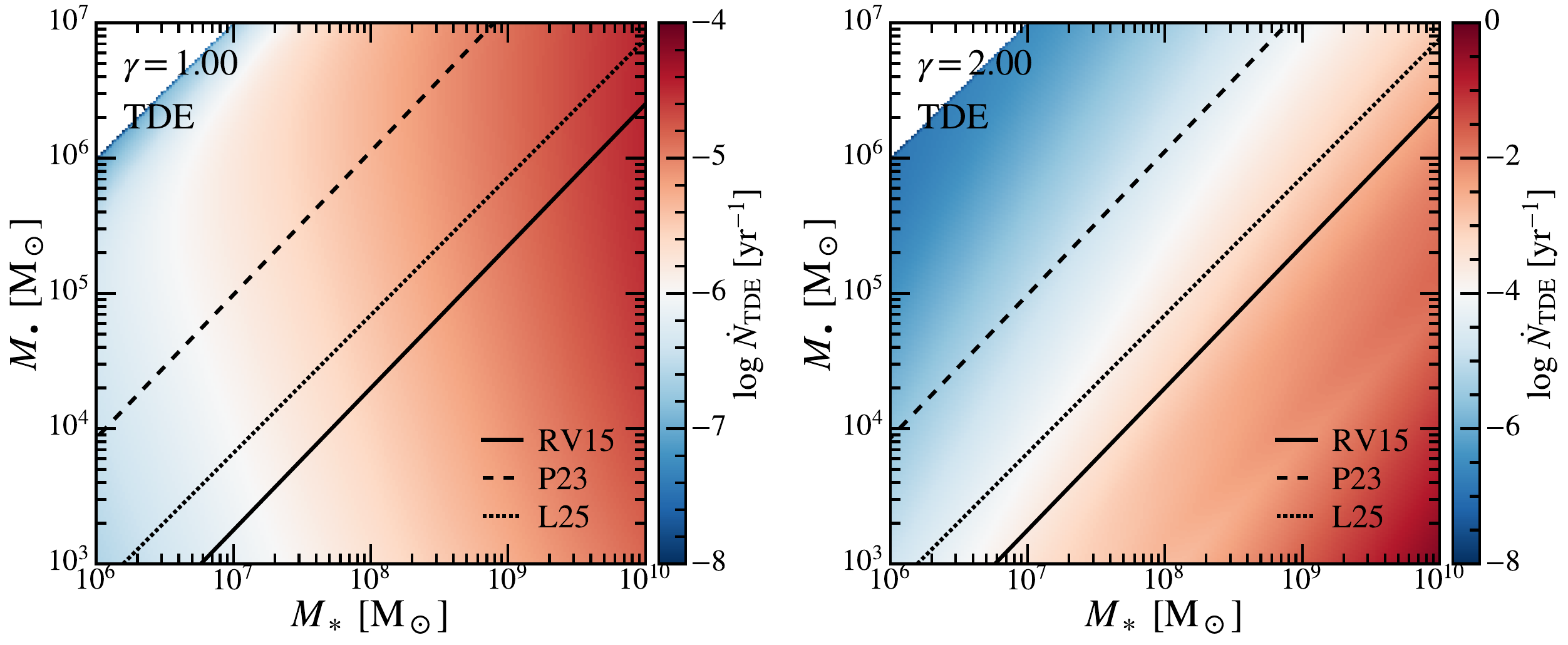}
    \caption{The rest-frame TDE rate of LRDs as a function of $M_\bullet$ and $M_\ast$, assuming the Kroupa PDMF with $m_\ast^{\rm max} = 2 \msun$ and that the underlying stellar distribution follows modified Dehnen profiles with $\gamma = 1$ (left) and $\gamma = 2$ (right). The RV15, P23, and L25 $M_\bullet \text{--} M_\ast$ relations are represented by solid, dashed, and dotted lines, respectively. For the $\gamma = 1$ case, the TDE rate increases with increasing BH mass in the lower BH mass regime, where the $\gamma = 1$ slope strongly influences the density profile. Even with the Bahcall--Wolf cusp, the density slope near the influence radius remains below the critical value of $1.42$. At larger BH masses, where the $-4$ outer slope has a stronger influence, as well as in the $\gamma = 2$ case, where the slope is always larger than $1.42$, the TDE rate decreases with increasing BH mass.}
    \label{fig:tde_rate}
\end{figure*}

\begin{figure*}
    \centering
    \includegraphics[width= 1 \linewidth]{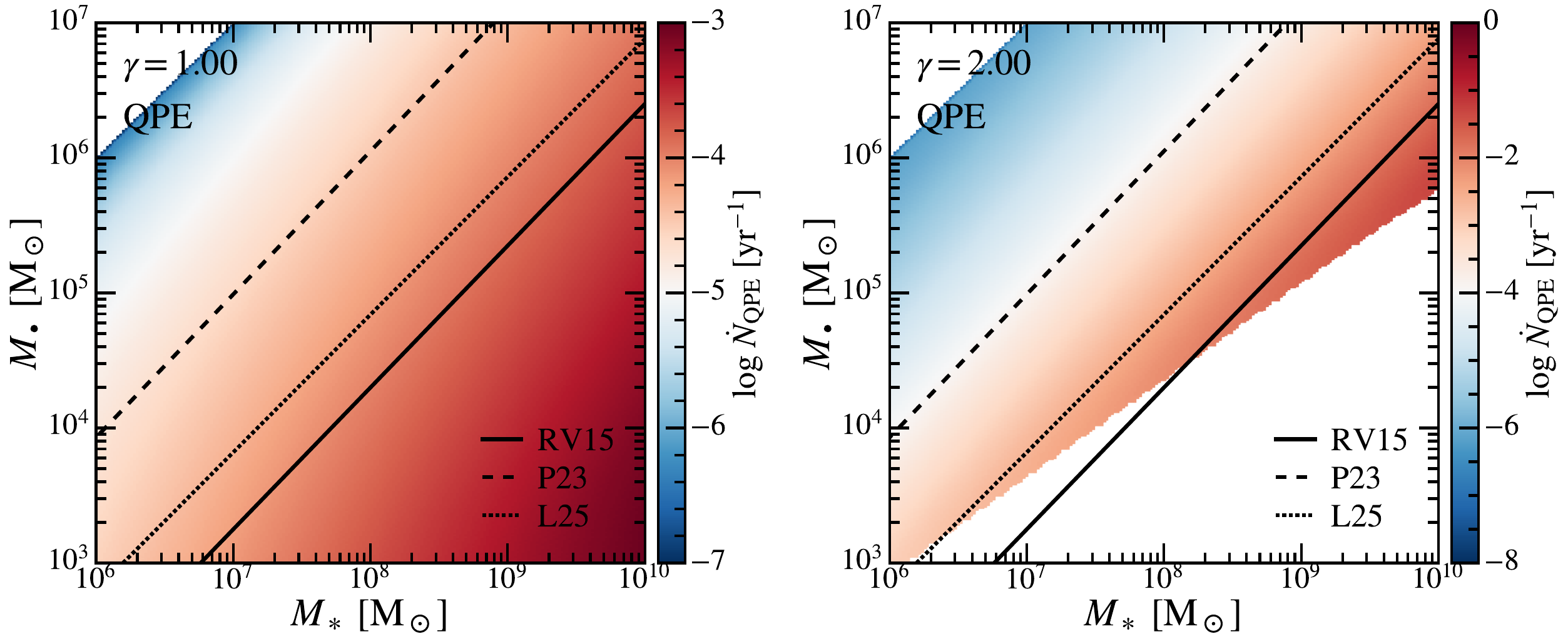}
    \caption{Similar to Figure \ref{fig:tde_rate}, but showing the rest-frame QPE rate of LRDs instead. The QPE rate broadly follows the same behavior as the TDE rate in the $\gamma = 2$ case, with the event rate decreasing with increasing BH mass. As a result of $r_{\rm ph}$ increasing the loss cone size by orders of magnitude, enhancements of $\sim 1 \text{--} 2\dex$ in the event rates relative to TDE are observed. However, in the $\gamma = 2$ case, the QPE rate calculation becomes invalid in the high-stellar mass, low-BH mass regime. This is because the more compact stellar distribution causes the influence radius $r_{\rm h}$ to rapidly approach $r_{\rm ph}$ as the BH mass decreases, resulting in large numerical errors, as discussed in Section \ref{ssec:diff}.}
    \label{fig:pee_rate}
\end{figure*}

% This arises because the large photosphere radius $r_{\rm ph}$ becomes comparable to the influence radius $r_{\rm h}$. Combined with the steeper $\gamma = 2$ profile, which reduces $r_{\rm h}$ more rapidly than $\gamma = 1$, this suppresses the enhancement seen in the $\gamma = 1$ case despite $r_{\rm ph}$ increasing the loss cone.

\begin{figure}
    \centering
    \includegraphics[width= \linewidth]{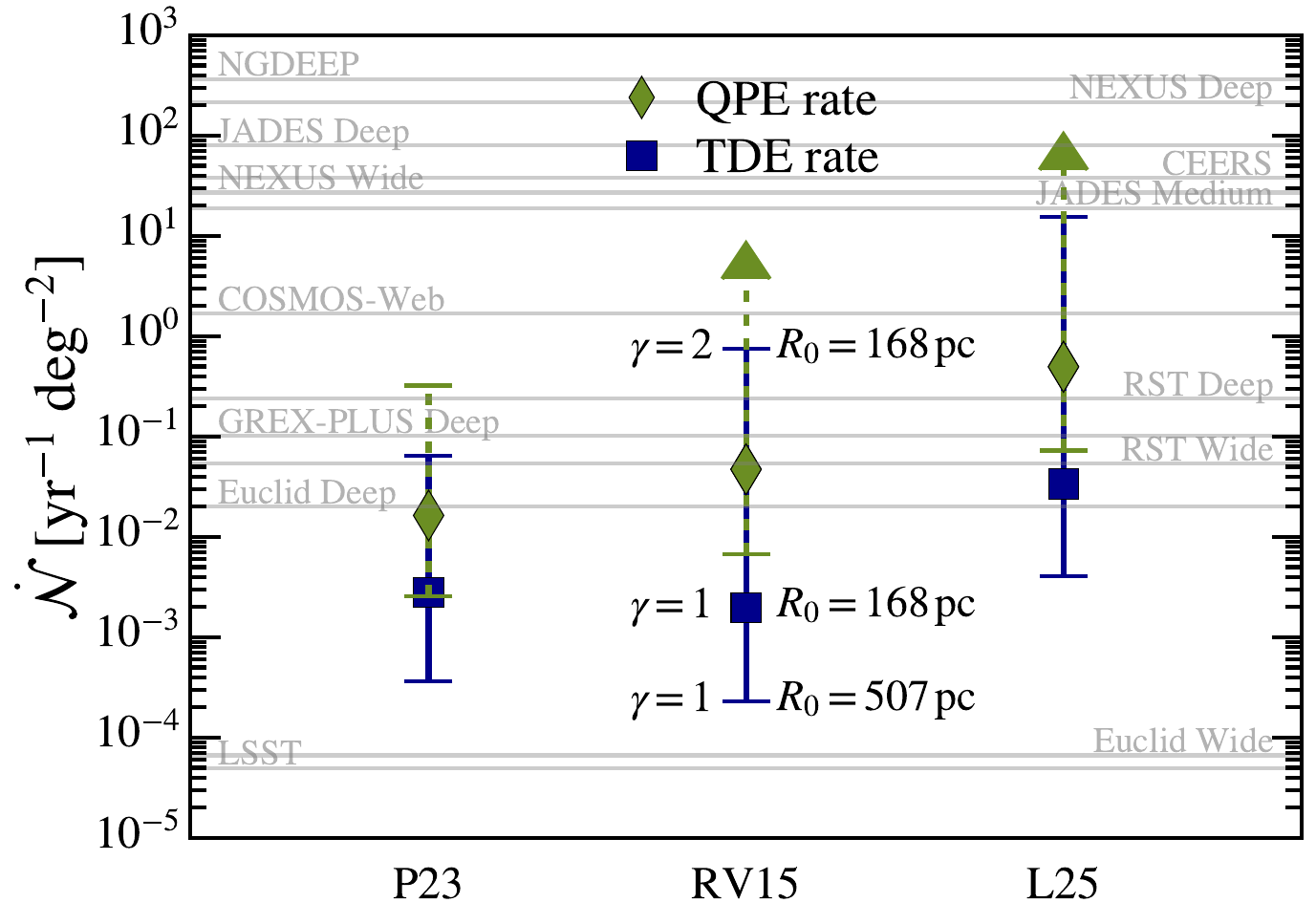}
    \caption{The per-degree square TDE (blue) and QPE (green) rates for LRDs across the three scenarios described in Section \ref{sec:result}, assuming a Kroupa PDMF with $m_\ast^{\rm max} = 2 \msun$. The upper error bars represent the $\gamma = 2, R_0 = 168\pc$ case, while the markers and lower error bars indicate the $\gamma = 1$ cases with $R_0 = 168\pc$ and $507\pc$, respectively. The upper bounds of the QPE rates in the RV15 and L25 scenarios represent only lower estimates, as most of the event rate in the lower-BH mass regime is unavailable, as shown in Figure \ref{fig:pee_rate}. Horizontal lines denote the inverse annual survey areas of various observational programs, i.e. the limit at which the experiments can detect a single event over the course of a year. The RV15 and P23 scenarios are generally difficult to distinguish, as variations in the intermediate density slope and effective radius introduce significant degeneracy. In contrast, the L25 scenario, which assumes a significant fraction of LRDs are obscured, predicts systematically higher TDE rates across all density slopes.}
    \label{fig:per_deg_rate}
\end{figure}

Figure \ref{fig:tde_rate} and \ref{fig:pee_rate} show the rest-frame per-galaxy TDE rate $\dot{N}_{\rm TDE}$ and the per-galaxy QPE rate $\dot{N}_{\rm QPE}$ as functions of $M_\bullet$ and $M_\ast$, computed for the modified Dehnen profile (Equation~\ref{eqn:dehnen_cusp}) with $\gamma = 1$ and $2$, assuming the size--mass relation in Equation \ref{eqn:size_mass_func}, as well as a Kroupa PDMF with $m_\ast^{\rm max} = 2 \msun$ (Equation \ref{eqn:Kroupa_pdmf}). In the lower-BH mass regime of the $\gamma = 1$ case, the TDE rate increases with increasing BH mass at fixed stellar mass. This is expected, as the event rate is dominated by the properties of the region near $r_{\rm h}$. In this regime, the density slope is intermediate between the Bahcall--Wolf $r^{-7/4}$ cusp and the outer profile; for the $\gamma = 1$ case, the value is $\simeq 1.37$, which remains below the critical value of $1.42$. At higher BH masses, the influence radius extends into the outer region, where the density slope approaches $4$. This transition also coincides with the onset of the Hill mass limit. For the $\gamma = 2$ case, in contrast, the density slope never falls below $7/4$, and the TDE rate therefore decreases by several orders of magnitude with increasing BH mass. The QPE rates across all stellar density profiles exhibit similar scaling behavior to this case, albeit with a $\sim 1 \text{--} 2 \dex$ increase in event rate, resulting from the larger loss cone associated with the larger photosphere radius $r_{\rm ph}$ compared to $r_{\rm t}$. In the $\gamma = 2$ case specifically, in the high-stellar mass, low-BH mass regime, contamination from numerical errors becomes impossible to mask out, and we therefore set the rate to zero, as discussed in Section \ref{ssec:diff}.

We also find that, when following the $M_\bullet\text{--}M_\ast$ relations, the TDE rate generally increases or remains stable with BH mass for all density profiles considered. This is in contrast to both theoretical predictions \citep[e.g.][]{Stone2016} and observational results \citep[e.g.][]{Yao2023} for the local galaxy population. This discrepancy is nevertheless understandable, as the systems examined in those studies typically possess steep cusps, with $\Gamma \sim 1$ ($\gamma \sim 2$), and are predominantly early-type galaxies. Such galaxies follow a much steeper size--mass relation, $a \sim M_\ast^{0.75}$, as shown by \cite{vdWel2014}. When we adopt this relation in our calculations, we recover the expected behavior, namely that the TDE rate decreases with $M_\bullet$ along the $M_\bullet\text{--}M_\ast$ relations. Additionally, compared to \cite{Inayoshi2024}, who find a per-galaxy TDE rate of $\sim 0.01 \yr^{-1}$, our calculation yields significantly lower rates. This discrepancy primarily arises from our assumption regarding the concentration of the stellar distribution in LRDs, which results in lower stellar densities and, consequently, lower TDE rates in our model.
%, i.e. that we adopt sizes comparable to high-redshift galaxies, whereas LRDs are often assumed to be much denser. The resulting difference in stellar density leads directly to the lower TDE rates in our model.

Figure \ref{fig:per_deg_rate} presents the per-degree square TDE and QPE rates for the three $M_\bullet\text{--}M_\ast$ scenarios under various assumptions for $\gamma$ and $R_0$ (Equations \ref{eqn:dehnen_cusp} and \ref{eqn:size_mass_func}). For our fiducial event rate, we adopt $\gamma = 1$ and $R_0 = 168\pc$. For the upper estimate, we consider the more compact profile with $\gamma = 2$, while at the lower-event rate end, we retain $\gamma = 1$ but increase the normalization of the size--mass relation to $R_0 = 507\pc$, corresponding to the median size of high-redshift galaxies. 

For TDEs, we find that, at fixed $R_0$, the differences between the first and second scenarios can be substantial, reaching up to an order of magnitude, depending on $\gamma$. Interestingly, the relative rates between these two scenarios reverse across the two regimes, with the RV15 case (local $M_\bullet\text{--}M_\ast$ relation) yielding a lower rate than P23 at $\gamma = 1$, but a higher rate at $\gamma = 2$. Increasing $R_0$ further suppresses the TDE rate for the $\gamma = 1$ RV15 case relative to P23. The third scenario produces considerably higher rates overall due to the larger intrinsic number density of LRDs. However, in all cases, because the TDE rate depends strongly on both $\gamma$ and $R_0$, it remains challenging to disentangle whether a given rate primarily reflects the underlying $M_\bullet\text{--}M_\ast$ relation or the assumed stellar density profile. For QPEs, the event rates are consistently higher than the TDE rates, even in the RV15 and L25 scenarios, where the higher estimates are likely substantially underestimated due to the missing values shown in Figure \ref{fig:pee_rate}. For $\gamma = 1$, the difference can approach $\sim 1.5\dex$ in the RV15 scenario. More importantly, the QPE rate is consistently higher in the RV15 scenario than in the P23 case, and higher in the L25 case than in RV15. As a result, combining TDE and QPE rate measurements may help more clearly distinguish between the different $M_\bullet\text{--}M_\ast$ scenarios. 

Another intriguing scenario is that LRDs may originate from supermassive stars \citep[SMSs; e.g.,][]{Nandal2026, Chisholm2026, Martins2026, Zwick2026, Naidu2026}. An upper limit of SMS mass is imposed by the general relativistic instability at $\sim 10^5\msun$~\citep[e.g.,][]{Woods2017, Nandal2024, Saio2024}. Assuming a stellar mass of $M_\ast = 10^8\msun$, the SMS mass of $M_{\rm SMS} = 10^{3 \text{--} 5}\msun$, and the same photosphere size as above, the expected observed QPE rates per LRD are $\sim 1.1 \text{--} 3.3 \times 10^{-5}\yr^{-1}$, with higher-mass SMSs yielding lower rates, increasing by roughly a factor of $2$ per decade decrease in SMS mass. This calculation assumes $\gamma = 1$ and $R_0 = 168\pc$. Per square degree, the expected number of QPEs over one year is $\dot{\mathcal{N}}_{\rm SMS} = 2.78 \text{--} 8.04 \times 10^{-2}\yr^{-1}\deg^{-2}$, which is comparable to the QPE rate in the RV15 scenario.

\subsection{Observational prospects}
\label{ssec:obs}

\begin{table*}
    \centering
    \addtolength{\tabcolsep}{19pt}
    \def\arraystretch{1.4}
    \begin{tabular}{l c c c c}
        \hline
        & \multicolumn{2}{c}{$N_{\rm TDE}$} & \multicolumn{2}{c}{$N_{\rm QPE}$} \\
        & $\gamma = 1$ & $\gamma = 2$ & $\gamma = 1$ & $\gamma = 2$ \\ [1ex]
        \hline\hline

        P23 (``Overmassive'') & $1.12 \times 10^{-3}$ & $2.59 \times 10^{-2}$ & $6.61 \times 10^{-3}$ & $0.130$ \\
        RV15 (``Local'') & $7.92 \times 10^{-4}$ & $0.302$ & $1.91 \times 10^{-2}$ & $>1.50$ \\
        L25  (``Tip of the iceberg'') & $1.36 \times 10^{-2}$ & $6.24$ & $0.201$ & $> 18.4$ \\
        \hline
    \end{tabular}
    \caption{The expected TDE and QPE event counts for different scenarios, assuming that 1000 LRDs are monitored over a period of 1 year. The normalization radius of the size--mass relation in Equation \ref{eqn:size_mass_func} is fixed at $R_0 = 168\pc$ for these calculations.}
    \label{tab:obj_rate}  
\end{table*}

Similar to \cite{Inayoshi2024}, we compare the predicted TDE and QPE rates with the inverse annual survey areas across various missions. These include several JWST programs: the Advanced Deep Extragalactic Survey \citep[JADES,][]{Eisenstein2026}, the CEERS survey \citep{Finkelstein2023}, the COSMOS-Web survey \citep{Casey2023}, the Next Generation Deep Extragalactic Exploratory Public survey \citep[NGDEEP,][]{Bagley2024}, and the North ecliptic pole EXtragalactic Unified Survey \citep[NEXUS,][]{Shen2024}. We also consider the Roman Space Telescope (RST) High Latitude Time Domain Survey for Supernova Cosmology across its deep and wide tiers \citep[e.g.][]{Rose2021}, the Galaxy Reionization EXplorer and PLanetary Universe Spectrometer \citep[GREX-PLUS,][]{Inoue2023}, \textit{Euclid} \citep[both the wide and deep surveys,][]{Laureijs2011,EuclidCollaboration2022}, and the LSST \citep[][]{Ivezic2019}. Due to the low expected number of TDEs, only wide-field surveys such as \textit{Euclid} and LSST would be capable of detecting them if LRDs possess shallower stellar density profiles. For these profiles, even though the QPE rates can be significantly higher than the TDE rates, they also remain too low to be detected by other surveys. In contrast, if LRDs exhibit steep cusps, most surveys (with the exception of JWST missions) would be able to detect TDEs and potentially discriminate between the scenarios, provided that some constraint on $\gamma$ and $R_0$ is established. If we instead take the approach of monitoring several objects rather than detecting transient events in a wide field, the expected events are still quite small. Table \ref{tab:obj_rate} shows the expected number of events if 1000 LRDs are monitored. Only in the L25 scenario with a steep cusp profile does the expected number of events substantially exceed 1 per year.

\section{Discussion \& Conclusion}
\label{sec:discussion}

In this paper, we investigate the TDE and QPE rates of LRDs using a dynamical framework. Calculations for QPE are performed by replacing the tidal disruption radius $r_{\rm t}$ with the photosphere radius $r_{\rm ph} = 1000 \AU$. We assume that the stellar distribution of LRDs follows the modified Dehnen family of density profiles with embedded Bahcall--Wolf cusps and varying intermediate density slopes. We adopt the size--mass relation of high-redshift galaxies lowered by $2\sigma$ in effective radius intrinsic scattering, motivated by the relative abundance of these populations. We consider the Kroupa PDMFs, with an upper mass cutoff of $m_\ast^{\rm max} = 2 \msun$, appropriate for a stellar population of age $T \sim 1 \Gyr$. The LRD comoving number density is obtained by integrating the bolometric luminosity function from \cite{Greene2026}. We compute the average TDE rate for LRDs across a total stellar mass range of $10^{7} \text{--} 10^{9} \msun$, weighting all masses within this interval equally. We explore three scenarios: (i) the $M_\bullet \text{--} M_\ast$ relation follows the apparent LRD relation from \cite{Pacucci2023}; (ii) it instead follows the local relation from \cite{ReinesVolonteri2015}; and (iii) the currently observed LRD population represents only a small, selection-biased subset of the intrinsic distribution, which is largely obscured.

We find that the three scenarios yield markedly different TDE and QPE rates for stellar distributions with steeper cusps ($\gamma \sim 2$). This difference is particularly pronounced in the third scenario, where the intrinsic number density is enhanced by approximately an order of magnitude. Interestingly, the ratio of TDE rates between the first and second scenarios reverses between the two extremes, $\gamma = 1$ and $\gamma = 2$. This reversal, together with the strong dependence of the event rate on the effective radius of LRDs, introduces a significant degeneracy between the $M_\bullet \text{--} M_\ast$ relation and the stellar distribution, making these scenarios difficult to disentangle observationally. In contrast, the QPE rate increases systematically for $M_\bullet \text{--} M_\ast$ relations that predict lower BH masses and remains significantly higher than the TDE rate of LRDs, even in cases where the QPE rate is substantially underestimated. 

%In cases where $r_{\rm ph}$ becomes comparable to the BH influence radius, the loss-cone calculation breaks down, leading to a sharp suppression of the event rate. This effect is more pronounced for cuspy profiles, where stars are more strongly concentrated in the inner regions.

Improved constraints on the stellar distribution, potentially enabled by higher-resolution observations of LRDs, would further aid in distinguishing between the three scenarios and in constraining the underlying $M_\bullet \text{--} M_\ast$ relation. Studies of high-redshift galaxies \citep[e.g.][]{Miller2025} indicate small S\'ersic indices of $n \sim 1 \text{--} 3$, which correspond closely to core-like profiles. However, these studies rely exclusively on S\'ersic profile fits, which have a limited ability to map inner density profile slopes of $\gamma < 1$. Consequently, the exact inner slope of the LRD stellar bulge remains uncertain. We also want to highlight an additional potential selection bias, the dependence of TDE duration on black hole mass. In the local population \citep[e.g.][]{Hammerstein2023,Yao2023}, TDE duration correlates positively with BH mass, meaning that events associated with more massive BHs may be more easily observed. This observational bias must be accounted for when considering the overall TDE rate. The low expected event rates for TDE and QPE are consistent with current constraints on the continuum variability of the majority of LRDs monitored \citep{Hayes2024,Kokubo2025,Tee2025,Zhang2025a,Liu2026}, and the reported long-term ($>100$ yr) variability in \citet{Zhang2025}, which may not be associated with the transient events discussed here. However, observing both types of events with upcoming wide-field surveys such as \textit{Euclid} and LSST could significantly increase the sample size and help better constrain the BH masses of LRDs.

%% Please use the acknowledgment and contribution environments. This will 
%% be anonomyized when the "anonymous" style option is used. 
\begin{acknowledgments}

We thank Dieu Nguyen, Kohei Inayoshi, and Yuhan Yao for useful discussions. We acknowledge the use of Google's Gemini and Anthropic's Claude for verifying the numerical implementation, as well as the use of OpenAI's ChatGPT for refining the text.
XS acknowledges support from NASA theory grant JWST-AR-04814.
AdG acknowledges support from a Clay Fellowship awarded by the Smithsonian Astrophysical Observatory.

\end{acknowledgments}

\begin{contribution}
%%This section gives authors the space to recognize author contributions. The text inside this environment is NOT counted towards the total word quanta. At a minimum, manuscripts are expected to include this text:

VT was responsible for the analysis, as well as writing and submitting the manuscript. 
XS developed the initial research concept, wrote the prototype of the analysis pipeline, and contributed to editing the manuscript.
OZ performed validation of the analysis and contributed to editing the manuscript.
AdG and RN helped refine the scientific idea and contributed to editing the manuscript.
MV secured funding for the project and contributed to editing the manuscript.

% All authors contributed equally to the paper.

%% But authors are expected to provide more specific details, e.g. 
%%
%%SC was responsible for writing and submitting the manuscript.
%%WWM came up with the initial research concept and edited the manuscript.
%%OTS obtained the funding and edited the manuscript.
%%EBF provided the formal analysis and validation. He also edited the manuscript.
%%GEH Supervised the undergraduates, wrote the software and administers the project github and Zenodo repositories.
%%
%% Authors can use the Contributor Role Taxonomy (CRediT) at
%% for ideas on how write a good statement tailored to their needs.

\end{contribution}

% \facilities{HST(STIS), Swift(XRT and UVOT), AAVSO, CTIO:1.3m, CTIO:1.5m, CXO}

\software{astropy \citep{Astropy2013,Astropy2018,Astropy2022},  
          numpy \citep{NumPy2020},
          scipy \citep{SciPy2020},
          }


\appendix

\section{Dimensionless loss-cone flux calculations}
\label{apd:dimless}

Taking the units of mass, length, and velocity as
\begin{equation}
    \label{eqn:units}
    [M] = M_\bullet, \hspace{0.5cm} [r] = r_{\rm h}, \hspace{0.5cm} [v] = \sqrt{G M_\bullet / r_{\rm h}},
\end{equation}
we calculate the dimensionless profiles
\begin{equation}
    \label{eqn:g_ast}
    \tilde{g} (\tilde{\epsilon}) = \frac{1}{\sqrt{8} \pi^2} \int_{0}^{\tilde{\epsilon}} \frac{{\rm d}^2\tilde{\rho}}{{\rm d}\tilde{\psi}^2} \frac{{\rm d}\tilde{\psi}}{\sqrt{\tilde{\epsilon} - \tilde{\psi}}}
\end{equation}
and
\begin{equation}
    \label{eqn:h_ast}
    \tilde{h} (\tilde{\epsilon}) = 3 \tilde{h}_{1/2} (\tilde{\epsilon}) - \tilde{h}_{3/2} (\tilde{\epsilon}) + 2 \tilde{h}_0 (\tilde{\epsilon}),
\end{equation}
where
\begin{align}
    \label{eqn:h_0_ast}
    \tilde{h}_{0} (\tilde{\epsilon}) &= \int_{\infty}^{\tilde{\epsilon}} \frac{{\rm d} \tilde{r}}{{\rm d} \tilde{\psi}} \frac{\tilde{r}^2 {\rm d} \tilde{\psi}^\prime}{\left(\tilde{\psi}^\prime - \tilde{\epsilon}\right)^{1/2}} \int_0^{\tilde{\epsilon}} \tilde{g}(\tilde{\epsilon}^\prime) {\rm d} \, \tilde{\epsilon}^\prime \\
    \label{eqn:h_1/2_ast}
    \tilde{h}_{n/2} (\tilde{\epsilon}) &= \int_{\infty}^{\tilde{\epsilon}} \frac{{\rm d} \tilde{r}}{{\rm d} \tilde{\psi}} \frac{\tilde{r}^2 {\rm d} \tilde{\psi}^\prime}{\left(\tilde{\psi}^\prime - \tilde{\epsilon}\right)^{\left(n+1\right)/2}} \int_{\tilde{\epsilon}}^{\tilde{\psi}^\prime} \left(\tilde{\psi}^\prime - \tilde{\epsilon}^\prime\right)^{n/2} \tilde{g}(\tilde{\epsilon}^\prime) {\rm d} \, \tilde{\epsilon}^\prime.
\end{align}
% We also calculate $\tilde{J}_{\rm c}^2 (\tilde{\epsilon})$ via interpolation of $\tilde{J}_{\rm c}^2 (\tilde{r})$ and $\tilde{\epsilon}_{\rm c} (\tilde{r})$, where $\tilde{\epsilon}_{\rm c} (\tilde{r})$ is the energy of the circular orbit with radius $\tilde{r}$. We take $\tilde{J}_{\rm lc}^2 (\tilde{\epsilon}) = 2 \tilde{r}_{\rm t}^2 (\tilde{r}_{\rm t}^{-1} - \tilde{\epsilon})$, using the $\psi(r_{\rm t}) \simeq G M_\bullet / r_{\rm t}$ approximation resulting from $r_{\rm t} \ll r_{\rm h}$. The ratio $q (\tilde{\epsilon})$ and loss cone flux $\mathcal{F} (\tilde{\epsilon})$ follow
% \begin{align}
%     \label{eqn:q_ast}
%     q (\tilde{\epsilon}) &= \frac{32 \pi^2}{3 \sqrt{2}} \ln{(\Lambda)} \frac{\langle m_\ast^2 \rangle}{M_\bullet \langle m_\ast \rangle} \left(\frac{r_{\rm t}}{r_{\rm h}}\right)^{-2} \frac{\tilde{h} (\tilde{\epsilon})}{\left(r_{\rm t} / r_{\rm h}\right)^{-1} - \tilde{\epsilon}}, \\
%     \label{eqn:F_ast}
%     \mathcal{F} (\tilde{\epsilon}) &= \frac{256 \pi^4}{3 \sqrt{2}} \frac{\ln{(\Lambda)}}{\ln{R_0^{-1}}} \frac{\langle m_\ast^2 \rangle}{\langle m_\ast \rangle^2} \sqrt{\frac{G M_\bullet}{r_{\rm h}^3}} \tilde{h} (\tilde{\epsilon}) \, \tilde{g} (\tilde{\epsilon}).
% \end{align}

%% For this sample we use BibTeX plus aasjournalv7.bst to generate the

\bibliography{sample701}{}
\bibliographystyle{aasjournalv7}



\end{document}